## Research Article

# Evolutionary-Based Sparse Regression for the Experimental Identification of Duffing Oscillator

**Saeideh Khatiry Goharoodi**  **,[1,2] Kevin Dekemele,[1] Mia Loccufier,[1] Luc Dupre,[1] and Guillaume Crevecoeur[1,2]**

[1]*Department of Electromechanical, Systems and Metal Engineering, Ghent University, B-9052 Zwijnaarde, Belgium*
[2]*EEDT Decision and Control, Flanders Make, Belgium*

Correspondence should be addressed to Saeideh Khatiry Goharoodi; saeideh.khatirygoharoodi@ugent.be

In this paper, an evolutionary-based sparse regression algorithm is proposed and applied onto experimental data collected from a Duffing oscillator setup and numerical simulation data. Our purpose is to identify the Coulomb friction terms as part of the ordinary differential equation of the system. Correct identification of this nonlinear system using sparse identification is hugely dependent on selecting the correct form of nonlinearity included in the function library. Consequently, in this work, the evolutionary-based sparse identification is replacing the need for user knowledge when constructing the library in sparse identification. Constructing the library based on the data-driven evolutionary approach is an effective way to extend the space of nonlinear functions, allowing for the sparse regression to be applied on an extensive space of functions. The results show that the method provides an effective algorithm for the purpose of unveiling the physical nature of the Duffing oscillator. In addition, the robustness of the identification algorithm is investigated for various levels of noise in simulation. The proposed method has possible applications to other nonlinear dynamic systems in mechatronics, robotics, and electronics.

## 1. Introduction

The Duffing oscillator is a nonlinear dynamic system with a considerable number of engineering applications and presents a key benchmark in nonlinear system analysis. The ordinary differential equation of this system consists of a cubic nonlinear term which can result in chaotic behavior and bifurcation. Suppression strategies are required to accommodate for this behavior in flexible robotic manipulators and high-precision mechatronic systems to increase their efficacy [1]. The control performance is however drastically affected by modeling errors in the system parameters [2, 3]. Also, the design of flexible manipulators and high-precision systems depend on the characteristics of the Duffing oscillator parameter variations [4]. Additionally, the design of harvesting devices from vibrations relies on the Duffing-type dynamic equations and when characterized well can be used as a tool for further analysis [5].

This work focuses on identifying the nonlinear ordinary differential equation of the Duffing system that consists of difficult to discover friction terms. Nonlinear system identification is a vast research field. The progress of this research area can be followed via several surveys including earlier works by Billings [6] and Mehra [7] as well as more recent studies [8–10].

In cases that the nonlinear model structure can be obtained from the first principles and is a priori known, the identification problem boils down to parameter estimation. Many works have been done in this area such as [11] where the physical parameter values are directly estimated using measured data. In many works such as [12–14], the least squares method is used in order to estimate the parameter values. Others report the usage of genetic programming for the same purpose [15, 16].

Parameter estimation using a fixed model structure based on captured data has been previously applied on Duffing oscillator-type systems. In [17], the parameters of a numerical fractional-order Duffing system have been identified using the sequential differential evolution method. Other algorithms such as nonlinear subspace identification method, particle swarm optimization, Volterra–Wiener-based model, and Wiener-type cascade model were used to numerically estimate the parameters of Duffing-type systems [18–21]. In a more recent attempt, authors in [22] have used a tailored sequential Monte Carlo algorithm within a Markov Chain Monte Carlo (MCMC) scheme to identify the parameters of Duffing in a Bayesian manner.

Alternatively, when the model structure is not a priori known, the form of the model needs to be discovered. Different black-box model structures can be considered to form the system equations. In [23], a modeling method for nonlinear systems using polynomial nonlinear state space equations was introduced. Furthermore, NARMAX models have been used in [24, 25] to represent nonlinear systems. Genetic algorithm and genetic programming have been also introduced in this field. In [26], genetic programming is used in a multiobjective fashion to generate global nonlinear models. Authors in [27] apply genetic programming to discover nonlinear differential equations. More examples of genetic algorithm application for system identification are [28–30]. Other modeling methods include but are not limited to neuro-fuzzy methods [31] and high-order neural network structures [32].

Black-box identification of the Duffing equation has also been a matter of investigation. In [33], explorative genetic programming is used to identify the model of a noisy Duffing–van der Pol oscillator using numerical simulation data. Artificial neural networks have been used to determine the mathematical model of the damped Duffing in [34]. A similar approach was proposed based on a set of basis functions and applying least-squares in [35].

A more exploitation-based nonlinear system identification approach was recently proposed in [36]. In this approach, a fixed matrix of candidate terms is first built upon prior expert knowledge. Subsequently, a linear system of equations is formulated using this matrix. The dominant terms in the constructed matrix later form the identified equation of the system. The sequential threshold least-squares algorithm is applied to find the true model of the system, depending on choosing the accurate value of the regularization parameter. A revised version of this method using the alternating direction method of multipliers (ADMM) has been successfully implemented on captured data from an experimental Duffing setup [37].

The biggest criticism towards sparse identification method lies in selecting *ad hoc* the appropriate library functions. This problem can be observed in [37] as the identification fails to discover the friction terms existing in the experimental data as these complex nonpolynomial terms are lacking in the library of functions. When identifying an experimental dataset, the friction forces within Duffing oscillators form an important model uncertainty that also arises in many other mechatronic applications such as in hydraulic actuators [38].

Nonlinear friction model parameters are reconstructed mostly based on a priori given friction model structures such as Coulomb and Stribeck friction models. Once these friction models are correctly identified, they can be used in control algorithms [39]. Consequently, in this paper, we aim at implementing an evolutionary-based sparse identification algorithm on the numerical and experimental Duffing system. The combination of genetic programming and sparse identification algorithm has been previously suggested in [40]; however, no methodology has been proposed so far.

In this paper, a revised version of sparse identification using the evolutionary-based sparse identification algorithm is studied for the first time to the authors' knowledge applied on a set of real-world experimental data. This paper is organized as follows. Section 2 describes the Duffing oscillator and the collected data from the setup and simulation. Section 3 provides details on the sparse regression algorithm. Section 4 briefly introduces the genetic programming method as the base for the evolutionary construction of the library and presents the evolutionary-based sparse identification algorithm to identify the model structure and parameters of the Duffing oscillator. Results and discussions are provided in Section 5, applying the identification method on experimental and numerical Duffing oscillator data. Conclusions are drawn in Section 6.

## 2. Problem Statement and Data Acquisition

In this work, the proposed algorithm is applied on the Duffing oscillator both numerically and experimentally. The cubic Duffing equation as a differential equation with a third-power nonlinear term is an example of a dynamic system that exhibits chaotic behavior and bifurcations. Experimental datasets are extracted from this setup to identify the underlying equation. Simulations from the same setup (using its characteristic parameters) implemented in the MATLAB environment [41], furthermore, allow to examine the efficiency of the algorithm and validating the presented method. The robustness of the algorithm is investigated on the data set by increasing added noise.

### 2.1. Duffing Oscillator: Theory and Experimental Realization

#### 2.1.1. Theoretical Description.
The mechanical Duffing oscillator with an imposed ground motion depicted as in Figure 1 is characterized by the following dynamic equation:

$$m\ddot{x} = -c(\dot{x} - \dot{z}) - k(x - z) - k_3(x - z)^3, \quad (1)$$

with $m$ being Duffing's mass, $c$ its linear damping, $k$ the linear stiffness, and $k_3$ the cubic stiffness.

A change of coordinates to the relative ground displacement $q \triangleq x - z$ yields

$$m\ddot{q} = -c\dot{q} - kq - k_3 q^3 - m\ddot{z}, \quad (2)$$

or the dynamics expressed in state space:

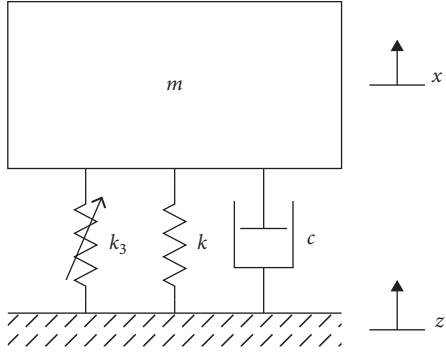

FIGURE 1: A mechanical Duffing oscillator subjected to imposed ground motion.

$$\begin{cases} \dot{q}_1 = q_2, \\ \dot{q}_2 = -\dfrac{c}{m}q_2 - \dfrac{k}{m}q_1 - \dfrac{k_3}{m}q_1^3 - \ddot{z}. \end{cases} \quad (3)$$

*2.1.2. Design Principle of Mechanical Duffing Oscillator.* To realize the mechanical Duffing oscillator, a mass-spring system, see Figure 2, is constrained to move along a designed track $y = f(x)$. The track's shape determines the linear and nonlinear stiffness, $k$ and $k_3$, (3).

If the mass is subject to a static force in the $x$-direction, the mass moves along the track until equilibrium is reached. It is now shown that the spring characteristic is nonlinear. The track exerts a reaction force on the followers attached to the spring, $R$, perpendicular to the track's curvature. The linear spring is compressed according to the track, imposing a force on the mass in the $y$-direction, $F_y = k_l y(x)$. The static applied force, the reaction force on the follower, and the reaction force on the mass are related by static equilibrium:

$$\begin{aligned} F_x &= 2R \sin(\theta), \\ F_y &= R \cos(\theta) \Longrightarrow F_x = 2F_y \tan(\theta), \end{aligned} \quad (4)$$

with $\theta$ being the tracks curvature's angle, related to the force profile by $\tan\theta = df(x)/dx$. If the track is $f(x) = ax^2 + b$, the spring characteristic is

$$F_x = 2k_l f(x)\dfrac{df(x)}{dx} = 4k_l a(bx + ax^3) = kx + k_3 x^3, \quad (5)$$

with $k = 4k_l ab$ and $k_3 = 4k_l a^2$. By machining a parabolic track, $f(x) = ax^2$, the linear coefficient can be simply tuned by shifting the profiles over a distance $b$.

*2.1.3. Experimental Setup.* The realized mechanical Duffing oscillator with the abovementioned design principle is shown in Figure 3(a). A mass with linear springs was fitted on a linear guide rail. Tracks with the shape with $a = 4\,\text{m}^{-1}$ were made from machined steel. The followers on the springs are SKF ball bearings. The linear springs have a stiffness of $k_l = 16.7\,\text{kN}$, according to the manufacturer, with the cubic stiffness then being $k_3 = 1.07\,\text{MN/m}^3$. The profiles can be shifted for adjusting the $b$ term in equation (5).

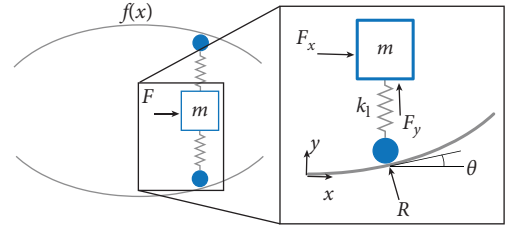

FIGURE 2: Design principle of the Duffing oscillator.

To impose the ground motion, the oscillator is put on a shaking table, here a Beckhoff linear permanent magnet motor. To measure Duffing's mass and shaking table displacement, accelerometer signals are integrated with the algorithm in [42]. For this algorithm to perform well, the signals should stay in a certain frequency band. The ground displacement imposed by the shaking table is limited in bandwidth by choosing a sine sweep and a random phase multisine.

The material contact between the followers and the track causes dry friction. The force in the $y$-direction $F_y$ will cause a perpendicular opposing friction force, $\mu F_y$, with $\mu$ the friction coefficient. The total opposing friction is

$$F_f = 2\mu k_l (ax^2 + b)\text{sgn}(v_x) = \mu_1 \text{sgn}(v_x) + \mu_2 x^2 \text{sgn}(v_x), \quad (6)$$

with $v_x$ the speed in the $x$-direction. Including the friction forces in the state space representation of the Duffing oscillator dynamics is

$$\begin{cases} \dot{q}_1 = q_2, \\ \dot{q}_2 = -\dfrac{c}{m}q_2 - \dfrac{k}{m}q_1 - \dfrac{k_3}{m}q_1^3 - \dfrac{\mu_1}{m}\text{sgn}(q_2) - \dfrac{\mu_2}{m}q_1^2\text{sgn}(q_2) - \ddot{z}, \end{cases} \quad (7)$$

in which the viscous damping $c$, linear stiffness $k$, and dry friction coefficients $\mu_1$ and $\mu_2$ have to be experimentally identified.

*2.1.4. Experimental Data.* The input of the dynamical equation (7), acceleration of the shaking table $\ddot{z}$, and the relative acceleration between the mass and shaking table $\ddot{q}$, captured from the described setup, are shown in Figures 4(a) and 4(b), respectively. The excitation signal of the experiment is a sine sweep from 2 to 20 Hz. The sampling time equals 0.488 ms.

**2.2. Duffing Oscillator: Numerical Data.** In order to validate the performance of the algorithm on numerical data, the described Duffing setup has been simulated in MATLAB. The state space model used for the purpose of simulation is the same as (3).

The control input of the system is a linear swept-frequency cosine presented in Figure 5(a). The noisy output acceleration obtained from numerical simulation is presented in Figure 5(b). The sampling time is 0.488 ms. The amplitude is selected such that the bifurcation is observable in the output. Considering both sets of experimental and numerical data in Figures 4 and 5, it can be noted that the bifurcation occurs sooner in simulation. The bifurcation of a Duffing oscillator occurs at a certain frequency of the input sweep. This frequency depends on the amplitude of the sweep, [43], which is different for the experiment and the simulation, explaining why the bifurcation happens at a different instant. The data for both (experimental and numerical) cases are divided in identification and validation parts.

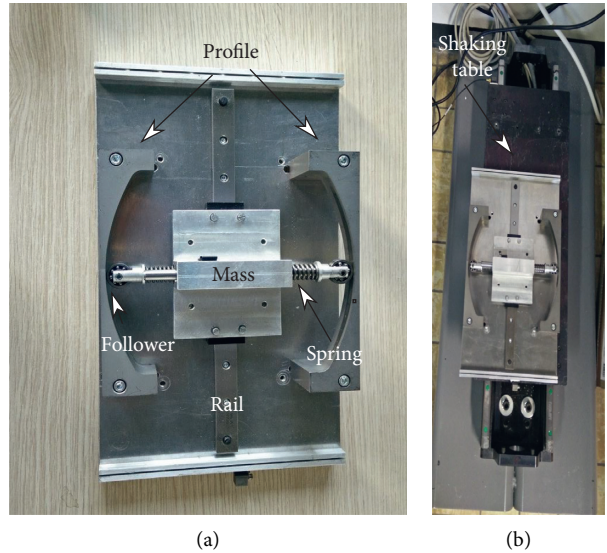

(a) (b)

Figure 3: (a) Realization of the Duffing oscillator design and (b) the Duffing oscillator on a shaking table.

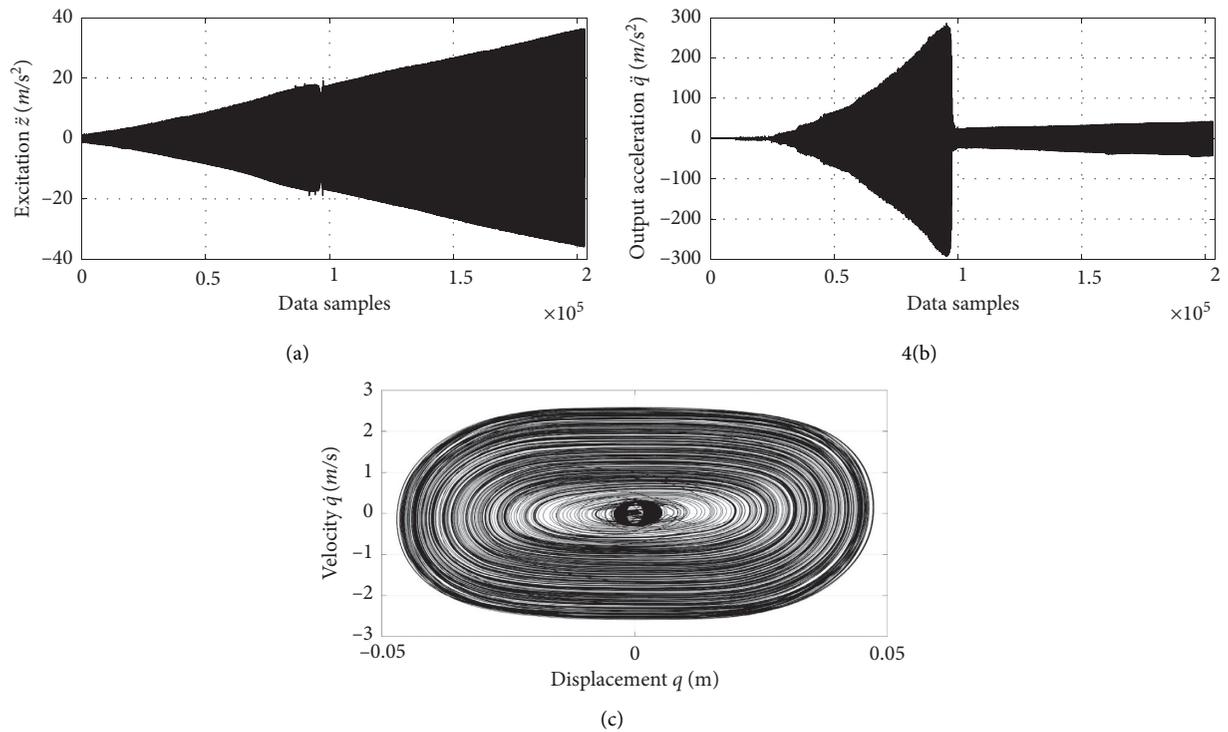

Figure 4: Experimental data: (a) the input, the acceleration of the shaking table $\ddot{z}$; (b) the output, the relative acceleration between the mass and shaking table $\ddot{q}$; (c) velocity versus displacement.

## 3. Sparse Regression

The aim of sparse regression in the field of system identification is to extract a low-dimension (sparse) representation of the system from a high-dimensional space of candidate representations using input and output data of the system. Considering $q \in \mathbb{R}^{n \times p}$ as the data matrix with $p$ state variables, each presented as a column of the matrix over $n$ time instants, sparse regression determines the state space equation as a general nonlinear function g:

$$\dot{q} = g(q, u), \quad (8)$$

where $u \in \mathbb{R}^{n \times 1}$ is the input of the system, $\dot{q} \in \mathbb{R}^{n \times p}$ is the time derivative of the states which can be measured or numerically calculated, and the q matrix (with derivatives $\dot{q}$) is assumed to be fully observable.

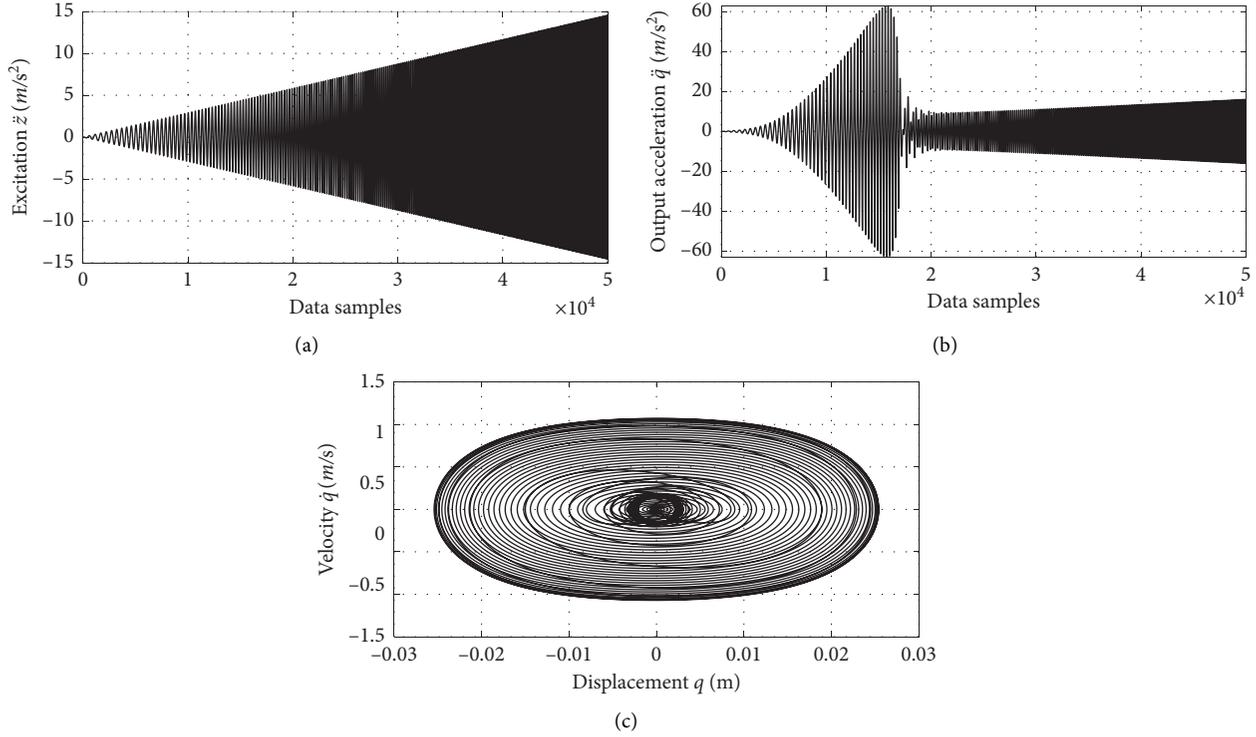

Figure 5: Numerical data. (a) The input, acceleration of the shaking table $\ddot{z}$. (b) The output, the relative acceleration between the mass and shaking table $\ddot{q}$. (c) Velocity versus displacement.

By introducing a library of terms as functions of the states and input of the system, the identification problem can be presented as finding the sparse matrix $\xi \in \mathbb{R}^{m \times p}$ [36]:

$$\dot{q} = A\xi, \quad (9)$$

where A is the library of (non)linear terms.

Choosing the right form of nonlinearity in the construction of the dictionary is essential in this approach which requires user knowledge. Equation (10) illustrates such a library. Each column, $m$, corresponds to a linear/nonlinear term as a function of the states or the input:

$$A(q, u) = \begin{pmatrix} | & | & | & | & | & | & | \\ 1 & q & q^2 & \cdots & u & u^2 & \cdots \\ | & | & | & | & | & | & | \end{pmatrix}_{n \times m}. \quad (10)$$

By solving equation (9), the dominant linear and nonlinear elements of the library A(q,u) will be chosen to combine linearly and form the equation of the system g in equation (8). The $\xi$ matrix is determined by minimizing a defined optimization problem. In this paper, we define the optimization problem as the elastic net regulator [44]:

$$\xi_{EN}^* = \arg\min_{\xi} \|A\xi - \dot{q}\|_2^2 + \lambda_1 \|\xi\|_1 + \lambda_2 \|\xi\|_2^2, \quad (11)$$

where $\lambda_1$ and $\lambda_2$ are the hyperparameters that are changed discretely. The order of magnitude for these hyperparameters is defined through the parameter sweep.

## 4. Evolutionary-Based Sparse Regression Methodology

*4.1. Genetic Programming.* Genetic programming (GP) is a subclass of genetic algorithms that was first presented by Koza in 1992 [45]. The basic idea of genetic programming is to evolve populations of equations based on the captured data and the fitness function evaluation of the simulation of each equation, where each equation is presented as a tree.

In the first generation, a population is randomly constructed by combining the numbers, variables, and mathematical operations. Terminal nodes of the trees are occupied by variables and numbers. The operators consisting of basic algebraic operations (+, −, ×, and /), functions (sin, cos, tan, abs, and sgn), or user-defined functions fill in the nonterminal nodes called the primitives. Afterwards, the population can vary in two ways: crossover and mutation. A crossover happens when two parent trees randomly exchange branches to form new offspring (Figure 6). Mutation involves random alteration of a parent's subtree (Figure 7). In the next step, the algorithm evaluates the fitness of each tree. The next generation is built based on the fitness evaluation. Following, the algorithm cycles through this loop until it reaches the stopping criteria or its convergence. A typical error metric such as least squares or root mean squared error is used as the fitness measure.

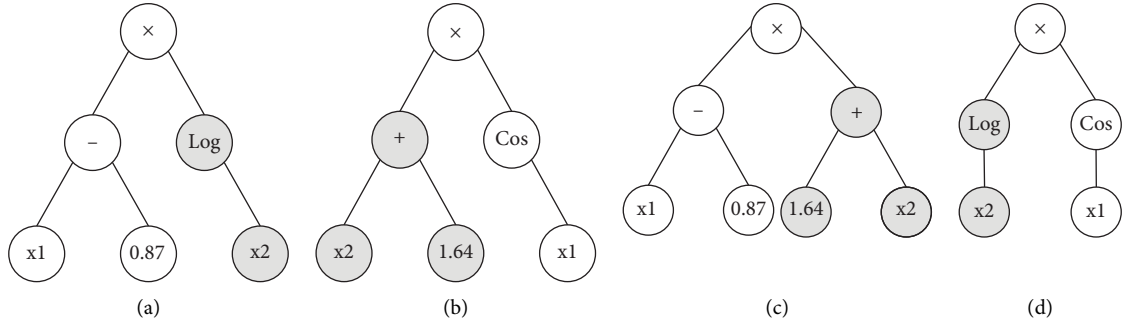

Figure 6: Illustration of the genetic programming crossover. (a) First parent before crossover with a randomly selected branch. (b) Second parent before crossover with a randomly selected branch. (c) First offspring after crossover. (d) Second offspring after crossover.

### 4.2. ESparse Algorithm.

Following the description of sparse regression and genetic programming, in this section, the proposed algorithm is described. As presented in Algorithm 1, the identification procedure consists of two main steps:

(1) Construction of the library ($\underset{n \times m}{A}$) using genetic programming
(2) Performing the sparse regression

In each iteration, an ODE equation is realized by solving a layered optimization problem: the individual trees in the population are used as the functions to build the $\underset{n \times m}{A}$ library in (10). Next, the sparse regression is performed on the constructed library by solving (11). Based on this approach alternative to a predefined library, the sparse regression is applied on a dynamic set of functions generated from the genetic programming. The advantage of this method is clearly its ability to generate an explorative library consisting of an extensive space of functions derived from the captured data. Moreover, this alternation between the explorative step 1 and the exploitative step 2 allows a reduction in the number of terms for regression.

## 5. Results and Discussion

In this section, the ability of identifying the correct form of the Duffing equation using the method from Section 4.2 in case of both numerical and noisy experimental datasets is analyzed. Both sets of data are captured from the Duffing oscillator described in Section 2, as a nonlinear dynamic system benchmark. In case of experimental data, we are specifically looking for the identification of the state space including the friction term as in equation (7). We also investigated the robustness of the algorithm with respect to noise in the data. By changing the level of added noise in simulation and how the accuracy of the identified model is affected by that noise provides a means to assess the robustness.

### 5.1. Numerical Duffing.

When applying the ESparse algorithm on the captured input/output dataset from Duffing oscillator simulations (Figure 5), the state space equation is identified. The first 16000 samples (the head of the arrow) are selected for validation, while the remainder are used for identification. For the numerical analysis to follow, the parameters in equation (3) are assumed to have the values $m = 0.49\,(\text{kg})$, $k = 487\,(\text{N}\cdot\text{m}^{-1})$, $k_3 = 1.07e6\,(\text{N}\cdot\text{m}^{-3})$, and $c = 1.8\,(\text{N}\cdot\text{s}\cdot\text{m}^{-1})$. The evolutionary parameters and values are presented in Table 1. Moreover, $q$, $\dot{q}$, and $\ddot{z}$ are the inputs of the GP denoted as X0, X1, and X2. The theoretical ODE equation together with the identified model for different levels of signal-to-noise ratio (SNR) is given in Table 2. The associated error percentage is calculated using validation data.

#### 5.1.1. Robustness Analysis.

To demonstrate the robustness of the algorithm, various levels of Gaussian white noise with zero mean were added to the data set. Figure 8 presents the tree of the identified equation in case of SNR = 19.5 dB. Moreover, the comparison between actual and identified validation data for SNR = 19.5 dB and SNR = 18.5 dB is presented in Figures 9(a) and 9(b), respectively.

A more general assessment for large ranges of signal-to-noise ratio (SNR) was performed, and the results are presented in Figure 10. Each data point in this figure corresponds to the mean of the accuracy percentage of 20 identification runs. Error bars are as well depicted that relate to the standard deviation. The results suggest that the proposed algorithm possesses the capability to reveal both the structure of the governing equation as well as the parameter values of the Duffing system. For SNR values from 20 to approximately 17 dB that correspond to increasing noise level, the accuracy of the identified parameter values decreases and ultimately the accuracy of the identification procedure itself. However, no additional terms appear in the discovered model, indicating the robustness of the presented algorithm. As can be observed in Figure 10, for low SNR (lower than approximately 17 dB) more terms are added to the equations. This clearly indicates

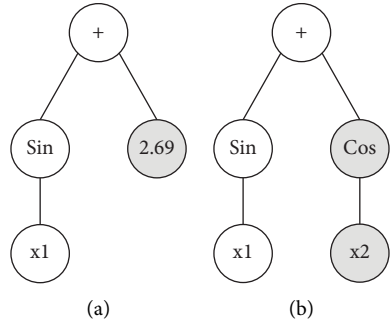

Figure 7: Illustration of the genetic programming mutation. (a) Parent before mutation with a randomly selected branch. (b) Offspring after mutation.

---

Require: time-varying measurement data: $\underset{n\times p}{q}$, $\underset{n\times p}{\dot{q}}$, and $\underset{n\times 1}{u}$. Population size $r$, number of generations $k$, and probabilities of crossover and mutation.
(1) Procedure:
(2)     Initialize the population of size $r$ randomly
(3)     For $i = 0: k$ do
(4)         Construct dictionary $\underset{n\times m}{A}$ based on the individuals
(5)         Solve the regression problem: $\xi_{EN}^{*} = \arg\underset{\xi}{\min} \|A\xi - \dot{q}\|_2^2 + \lambda_1 \|\xi\|_1 + \lambda_2 \|\xi\|_2^2$
(6)         Compute the fitness function: mean square error
(7)         Generate new population using crossover and mutation
(8)     End for
(9) End procedure

Algorithm 1: ESparse algorithm.

---

Table 1: Evolutionary parameters for the numerical Duffing.

| Evolutionary parameter | Value |
| --- | --- |
| Population size | 80 |
| Crossover rate | 0.9 |
| Mutation rate | 0.1 |
| Number of generations | 30 |
| Basis functions | Plus, minus, times, abs, sgn |

that the data become overfitted by the identified model ultimately resulting in deteriorated accuracies.

### 5.2. Experimental Duffing.
Similar to the numerical Duffing data, ESparse is applied onto experimental Duffing data with the purpose to identify the Duffing equation. We conducted three experiments with the same input acceleration profile under the same conditions, resulting in data presented as in Figure 11. In all cases, the first 90000 data samples of the control input and the output are selected for validation and the rest are used for training.

The evolutionary parameters of the genetic programming are given in Table 3. Table 4 summarizes the identified model obtained from the three experiments using the ESparse algorithm. Additionally, Figure 12 presents the tree of the identified equation with 5.6% error. The terms appearing in these equations are well supported by the theoretical model from equation (7) that includes Coulomb friction. The results demonstrate the ability of the algorithm to identify nonpolynomial nonlinearities. Comparison between the actual and identified output acceleration data is illustrated in Figure 13. A clear correlation between the two sets of data can be observed from Figure 13(b).

### 5.3. Comparison with Other Available Methods.
To substantiate the advantages of the proposed ESparse algorithm, a comparison with other available methods with respect to performance measures run time and % error are drawn in Table 5. Sparse regression (see Section 3) and genetic programming (see Section 4.1) are applied on the same dataset. For the purpose of having a fair comparison, crossover and mutation probabilities as well as the employed basis functions applied for genetic programming are the same as those employed in the ESparse algorithm (Table 3). However, to achieve the correct model of the system using genetic programming, the population size and number of generations have to increase to 250 and 80, respectively. As suggested by the results, the ESparse algorithm is capable of converging to the model with the same level of accuracy with much less computational effort. As for the sparse regression method, we had to manually include the sign function as being part of the library of (non)linear terms (coming from knowledge gained with the ESparse algorithm) since otherwise the model structure cannot be discovered, whereas the ESparse algorithm automatically builds the proper library using genetic programming.

Table 2: Identified models by the ESparse algorithm, numerical Duffing.

| | Identified ODE equation | % error |
|---|---|---|
| Theoretical reference | $\ddot{q} = -3.67\dot{q} - 993.88q - 2.18e6q^3 - \ddot{z}$ | - |
| SNR = 20 dB | $\ddot{q} = -3.67\dot{q} - 994.19q - 2.18e6q^3 - \ddot{z}$ | 0.7 |
| SNR = 19.5 dB | $\ddot{q} = -3.67\dot{q} - 994.81q - 2.19e6q^3 - 0.99\ddot{z}$ | 1.9 |
| SNR = 19 dB | $\ddot{q} = -3.67\dot{q} - 984.64q - 2.15e6q^3 - \ddot{z}$ | 3.7 |
| SNR = 18.5 dB | $\ddot{q} = -3.73\dot{q} - 867.83q - 1.72e6q^3 - 1.02\ddot{z}$ | 11.2 |

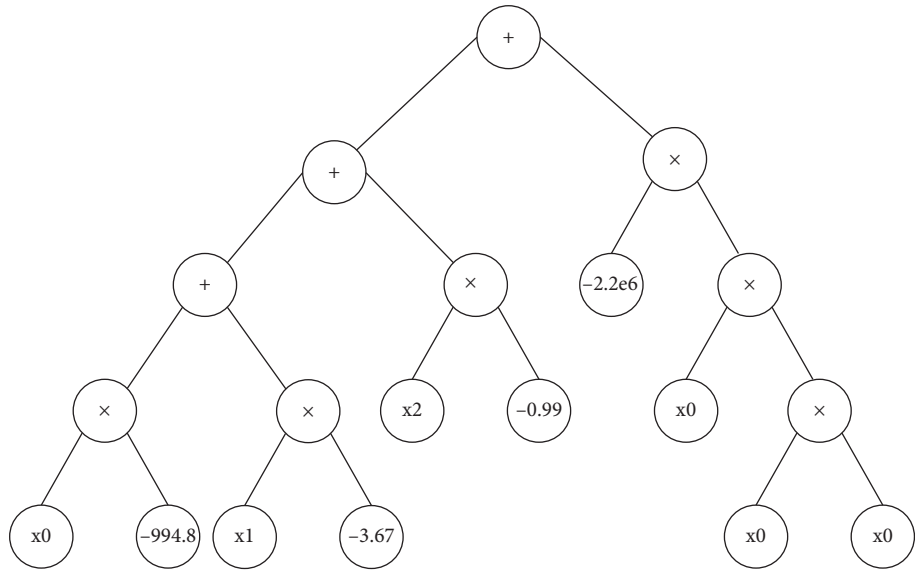

Figure 8: Tree presentation of the identified numerical Duffing with SNR = 19.5 dB.

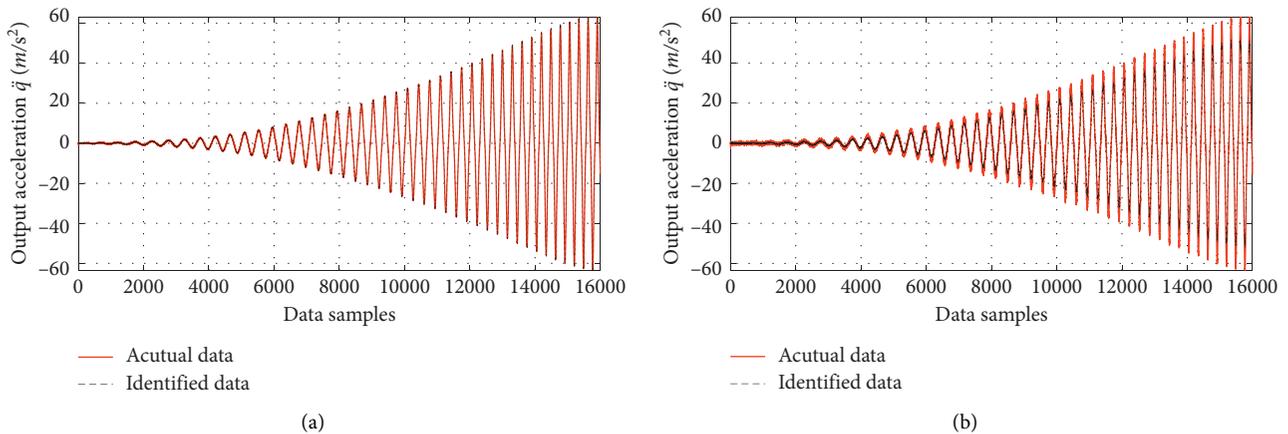

Figure 9: Comparison of the actual and identified numerical Duffing for (a) SNR = 19.5 dB and (b) SNR = 18.5 dB.

### 5.4. Advantages and Limitations of the Method

#### 5.4.1. Advantages.
The proposed methodology has major benefits in comparison with sparse regression and genetic programming-based methods for nonparametric identification. The evolutionary-based sparse regression requires lower computational effort relative to genetic programming-based algorithms. For GP-based algorithms to converge to the true solution, large populations with high number of generations are typically required. Nonetheless, the presented ESparse algorithm has the ability to converge to the correct model with less computational effort and having a balanced model complexity since ESparse alternates between exploration (genetic programming) and exploitation (sparse regression). Therefore, the algorithm can discover the system equation with fewer generations and smaller populations.

As for the sparse regression method, the strict model assumptions prior to identification can limit the model complexity while the dynamic library of functions in evolutionary-based sparse regression allows for discovery of more complex models by extending the search space and replacing the need for user knowledge for the construction of the library with data-driven GP step.

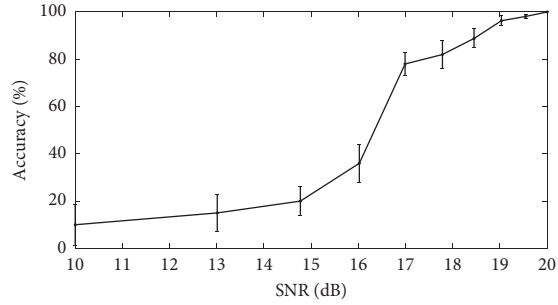

Figure 10: Mean and standard deviation of the identification accuracy for various levels of SNR.

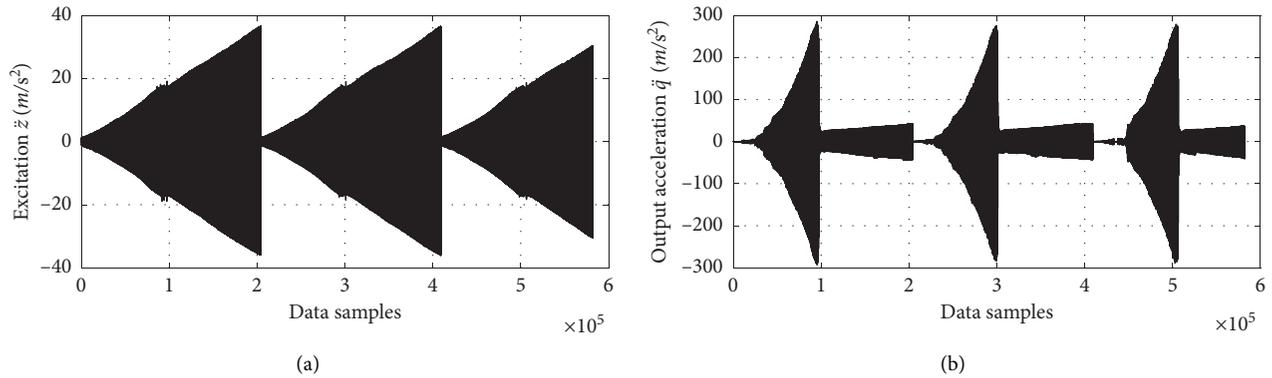

(a)

(b)

Figure 11: Experimental data. (a) The input, the acceleration of the shaking table $\ddot{z}$. (b) The output, the relative acceleration between the mass and shaking table $\ddot{q}$.

Table 3: Evolutionary parameters for the experimental Duffing.

| Evolutionary parameter | Value |
| --- | --- |
| Population size | 150 |
| Crossover rate | 0.8 |
| Mutation rate | 0.2 |
| Number of generations | 40 |
| Basis functions | Plus, minus, divide, times, abs, sgn |

Table 4: Identified models by the ESparse algorithm, noisy experimental Duffing.

| Exp. | Identified Duffing | % error |
| --- | --- | --- |
| 1: | $\ddot{q} = -1.10\dot{q} - 691.92q - 2.37e6q^3 - 2.93\mathrm{sgn}\dot{q} - 7.63e3q^2\mathrm{sgn}(\dot{q}) - 1.02\ddot{z}$ | 3.9 |
| 2: | $\ddot{q} = -1.23\dot{q} - 714.60q - 2.24e6q^3 - 4.11\mathrm{sgn}\dot{q} - 8.23e3q^2\mathrm{sgn}(\dot{q}) - 1.04\ddot{z}$ | 5.6 |
| 3: | $\ddot{q} = -1.22\dot{q} - 716.11q - 2.35e6q^3 - 3.81\mathrm{sgn}\dot{q} - 8.26e3q^2\mathrm{sgn}(\dot{q}) - 1.04\ddot{z}$ | 4.7 |

*5.4.2. Limitations.* Although the proposed algorithm allows to identify more complex nonpolynomial terms in the equation such as friction terms, the basic building blocks are required to be included in the pool of the basic functions of the GP algorithm. Otherwise, the identified system will only be composed of available blocks which may not represent the nature of the system accurately.

## 6. Conclusion

In this paper, an evolutionary-based sparse regression algorithm for discovering both the structure and the parameter values of the system has been proposed. The methodology is used for the purpose of identifying the Duffing oscillator system using both numerical and noisy experimental data. In case of numerical Duffing, the data are polluted with different levels of noise to study the robustness of the algorithm. Furthermore, the approach is challenged to discover governing dynamics that include nonpolynomial nonlinear Coulomb friction terms, from noisy experimental Duffing data. As shown by the percentage of the identification error, the algorithm is effective in unveiling the physical nature of the Duffing oscillator. The proposed method has possible applications to other nonlinear systems such as in mechatronics, robotics, and electronics.

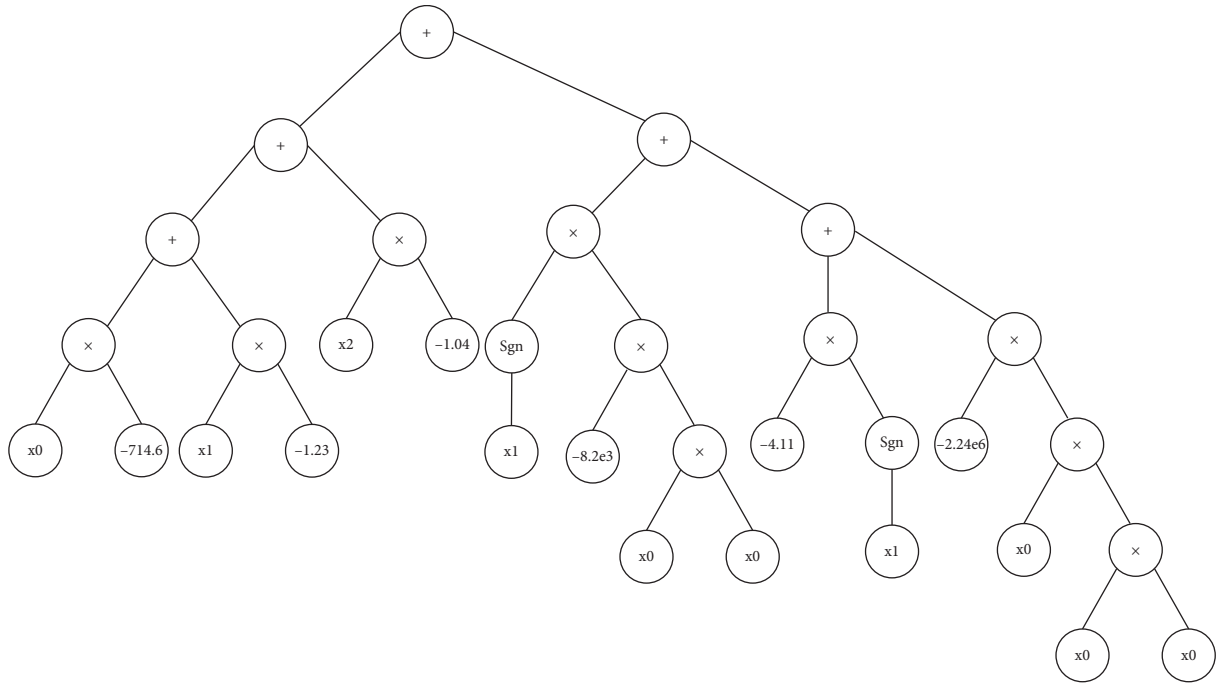

Figure 12: Tree presentation of the identified experimental Duffing 5.6% error.

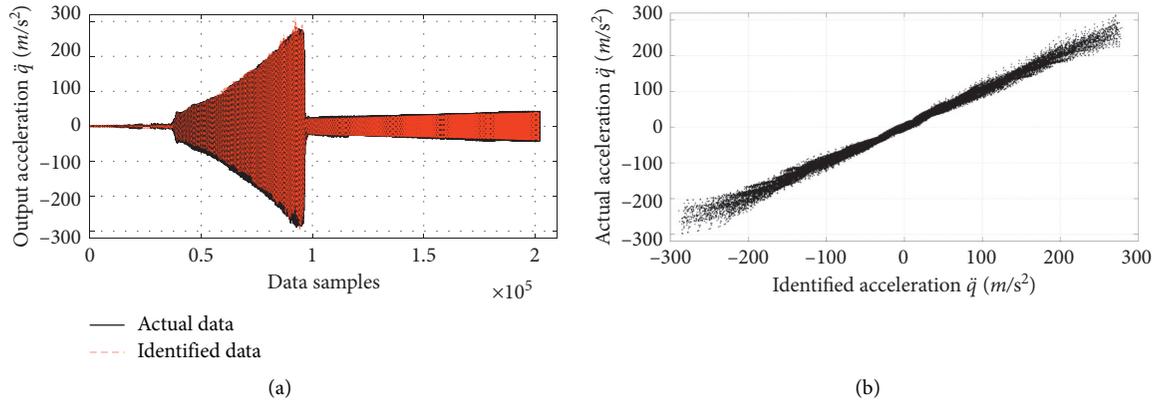

Figure 13: Comparison of the actual and identified experimental Duffing: (a) comparison over time and (b) actual acceleration versus identified.

Table 5: Comparison of performance measures, experimental Duffing.

| Method | Run time (s) | % error |
| --- | --- | --- |
| Genetic programming (exp. 1) | 449.411 | 5.7 |
| Genetic programming (exp. 2) | 425.884 | 5.3 |
| Genetic programming (exp. 3) | 539.042 | 4.1 |
| Sparse regression (exp. 1) | 2.092 | 3.9 |
| Sparse regression (exp. 2) | 2.563 | 4.6 |
| Sparse regression (exp. 3) | 2.677 | 5.2 |
| ESparse algorithm (exp. 1) | 12.626 | 3.9 |
| ESparse algorithm (exp. 2) | 12.171 | 5.6 |
| ESparse algorithm (exp. 3) | 12.125 | 4.7 |

## Data Availability

All the data are included in the article. If there is a further demand for data, the author can provide depending on their availability.

## Conflicts of Interest

The authors declare that there are no conflicts of interest regarding the publication of this paper.

## Acknowledgments

This work was supported by the ICON project Multi-Sensor and MODA of Flanders Make, the Strategic Research Centre for the Manufacturing Industry, and the FWO research project G.0D93.16N. This research received funding from the Flemish Government under the "Onderzoeksprogramma Artificiële Intelligentie (AI) Vlaanderen" Programme.